\documentclass[]{spie}  

\usepackage{amsmath,amsfonts,amssymb}
\usepackage{graphicx}
\usepackage{subcaption}
\usepackage[colorlinks=true, allcolors=blue]{hyperref}

\title{Energy Recovery System for Large Telescopes}

\author[a]{Aleksej Kiselev}
\author[a]{Matthias Reichert}
\author[b]{Tony Mroczkowski}

\affil[a]{OHB Digital Connect, Weberstra\ss e 21, D-55130 Mainz, Germany}
\affil[b]{European Southern Observatory, Karl-Schwarzschild-Str.\ 2, Garching 85748, Germany}

\authorinfo{Further author information: (Send correspondence to A.K.)\\
A.K.: E-mail: aleksej.kiselev@ohb.de\\
M.R.: E-mail: matthias.reichert@ohb.de\\
T.M.: E-mail: tony.mroczkowski@eso.org}

\begin{document} 
\maketitle

\begin{abstract}
In this paper, a kinetic energy recovery system for large telescopes is presented, with the Atacama Large Aperture Submm Telescope (AtLAST) as a possible target application. The system consists of supercapacitors integrated in the DC-link of motor inverters through a bidirectional DC-DC converter. The optimal system design, based on the energy flow analysis within the telescope's power electronics, is introduced. The proposed system is simulated as part of the telescope's drives, providing not only a significant reduction in energy consumption of the telescope due to motion, but also remarkably reducing (or shaving) grid power peaks. We find that the system presented here can contribute to making both current and future observatories more sustainable.
\end{abstract}


\section{INTRODUCTION}
\label{sec:intro}  

Large telescopes constitute massive structures that must be accelerated and decelerated quickly and often nearly continuously throughout operations.
In fact, large aperture telescopes are among the most massive steerable structures on land.
Thus, a large portion of their energy and power budgets are dedicated to mount motion.  Further, such facilities are often sited far from the local power grid, meaning energy must be considered more carefully as a commodity.
For instance, the Atacama Large Aperture Submm Telescope (AtLAST)\footnote{\href{https://atlast-telescope.org/}{https://atlast-telescope.org/}} aims to be a 50-meter telescope observing at frequencies 30-950~GHz, situated in the Atacama Desert in Chile at an elevation of $\approx$ 5050 meters above sea level \cite{2024arXiv240218645M,2024arXiv240608611R}. It is planned to be fully powered off-grid by a renewable power system \cite{Viole2023}. 

AtLAST will be a wide field of view telescope, able to perform fast scans \cite{2022PhRvD.105d2004M, 2024arXiv240210731V} with velocities up to $3^\circ \rm s^{-1}$ and accelerations up to $1^\circ \rm s^{-2}$.
This results in a very high motion power demand. On the other hand, deceleration inevitably follows acceleration, during which the telescope's electric motors inherently generate electric energy. Nonetheless, this energy is dissipated into heat by braking resistors in every major telescope we are aware of. The reason is that most fast-scanning telescopes (where the benefit of the energy recovery is higher) are smaller, so the dissipated energy is small compared to the overall energy budget. However, in the light of climate change, wasting this energy is increasingly difficult to justify \cite{Valenzuela-Venegas2023}. Additionally, and especially important from an economic perspective, is the possibility of substantial reduction in the size of the necessary power grid infrastructure that will be achieved through the reuse of motion energy. 

Energy recovery has been implemented in various forms since the early days of using electric motors for motion, e.g., in the first electric vehicles and trains \cite{History_ReGen} in the last century. There are various approaches of realization for different applications. In this paper we detail the application of an energy recovery scheme based on the system has been proposed for elevators \cite{Jabbour} to large telescopes. The system consists of supercapacitors (supercaps) to store the recuperated energy, and a bidirectional DC-DC converter to gain full control of the supercaps' charge and discharge cycles. The proposed system not only improves the efficiency of the telescope's drives significantly by reusing the deceleration energy during acceleration, but also distributes grid power peaks over time as a low but constant grid load, which allows a downsizing of the required power infrastructure. The system concept is applied to AtLAST as an ideal representative of a large telescope with a very high motion power demand. 
However, we note that while the gains in energy efficiency and power infrastructure downsizing are most dramatic for fast scanning and survey telescopes (e.g.\ the Vera Rubin Observatory\cite{2017arXiv170804058L},  Simons Observatory\cite{SO2019}, and CMB-S4 \cite{Abazajian2019}), the system we propose is more generally applicable to many facilities operating across the electromagnetic spectrum.

This paper is organized as follows. First, in Sec.~\ref{sec:Energy_Flow} the energy flow in telescope's main drives is analyzed and the scheme of the energy recovery system introduced. In Sec.~\ref{sec:Motion_Scenarios}, a Lissajous daisy scan is presented as one possible representative motion scenario for a telescope like AtLAST. Next, in Sec.~\ref{sec:SystemDesign} the optimal design of the proposed energy recovery system is described. Finally, simulation results of the system in the case of application to AtLAST are presented and discussed in Sec.~\ref{sec:SimResults}.

\section{Energy Flow in Telescope's Main Drives}\label{sec:Energy_Flow}

\subsection{Classic Power Electronics Scheme}\label{sec:ClassicPowerElectronics}
\begin{figure}[ht]
    \centering
    \includegraphics[width=1\textwidth]{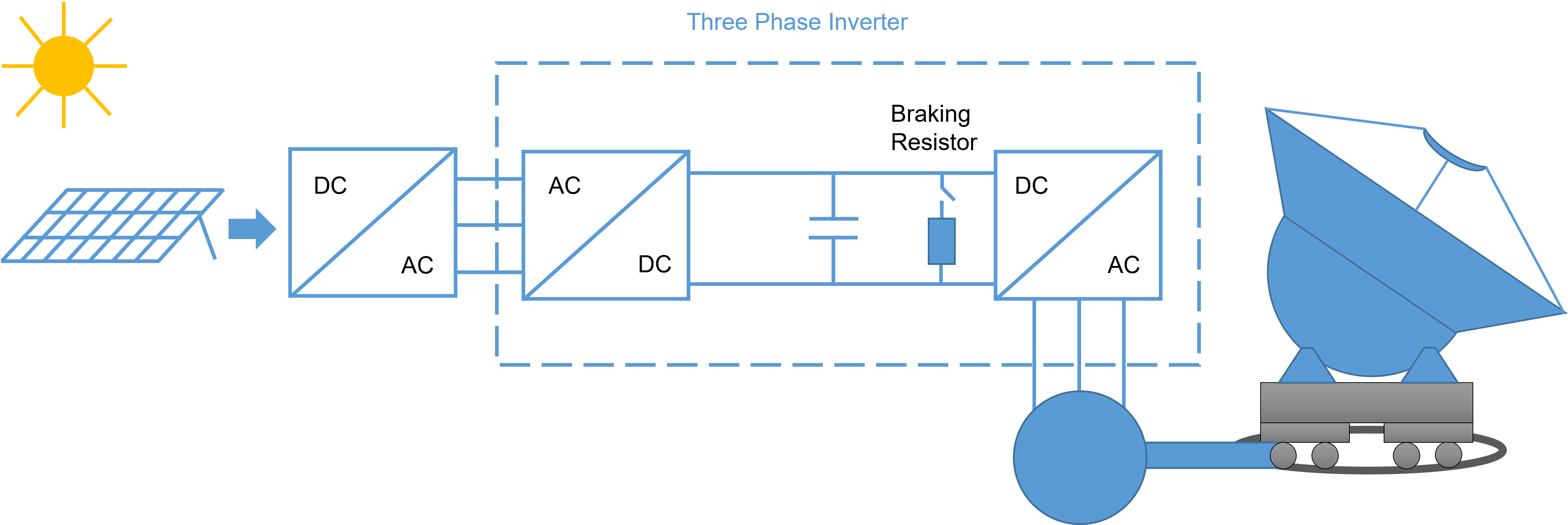}\vspace{0.2cm}
    \caption{Classic power electronics scheme of a telescope's axis with AtLAST as an example.}
    \label{fig:PowerElectronicsAZ}
\end{figure}

Typically, large telescopes have two steerable axes, i.e., the horizontal azimuth axis (AZ-axis) and the vertical elevation axis (EL-axis). The axes movement is realized by electric motors with corresponding power electronics. In the past, DC motors were mostly used. However, nowadays three-phase motors (induction machines or synchronous motors) are the state-of-the-art, since they have no brushes and thus, no wearing. To power an induction machine or synchronous motor, a three-phase inverter is required. It basically consists of three main components: an AC/DC converter to rectify the AC voltage from the grid to the DC voltage, a DC-link capacitor to smooth the rectified voltage and a subsequent DC/AC converter to provide AC voltage with the proper frequency to the motor. The simplified scheme of the AZ-axis is shown in Fig.~\ref{fig:PowerElectronicsAZ}. 

Generally, the mechanical (kinetic) power of an axis $P_m$ is defined as:
\begin{equation}
    P_m = \omega_m T_m
\end{equation}
where $T_m$ is the torque  of the axis and $\omega_m$ the corresponding angular velocity. Since the velocity and torque have the same sign during acceleration (either positive or negative), the power is positive, resulting into an electrical energy flow from the grid to the motors. From the electric perspective, in this scenario the telescope is a load. 

During deceleration, the sign of velocity and torque are opposite, and the axis generates power. Thus, kinetic energy of the axis is converted by the motors into electrical energy, flowing into the DC-link capacitor of the inverter. Since the rectifier part of an inverter is typically unidirectional, the generated energy cannot be fed back to the grid. That makes a braking resistor in parallel to the DC-link capacitor essential. When the DC-link voltage exceeds a maximum value, the braking resistor will be switched on and the DC-link capacitor discharged over it. Instead of reusing this energy for the next acceleration phase, it is converted into heat, with a significantly increased energy consumption and decreased efficiency of the drive system as a result.

\subsection{Proposed Power Electronics Scheme}
There are two principle solutions for the problem described above. The first is to use bidirectional inverters, which allow one to feed the recuperated energy back to the grid. The second solution is to integrate an additional energy storage in the DC-link to be able to take the total recuperated energy, without overloading the DC-link capacitor. The first solution is simpler from the technical perspective, however the second approach has a significant advantage: power peaks shaving. Since the recuperated energy stays within the drive's power electronics, it serves a substantial energy part required for the following acceleration. As result, the grid load is significantly reduced, which can be considered during the grid infrastructure planning, for example, by installing a smaller transformer. Because of this crucial advantage we clearly prefer the DC-link based energy storage and present in this paper its technical realization. 

For the electric energy storage, two main of-the-shelf techniques are theoretically possible: batteries or supercapacitors (supercaps). Batteries have a higher energy density but a significantly lower power density, compared to supercaps \cite{Andreev}. The telescope's acceleration and deceleration, on the other hand, is typically associated with a high but short power peak. That makes supercaps an ideal solution. 

\begin{figure}[ht]
    \centering
    \includegraphics[width=1\textwidth]{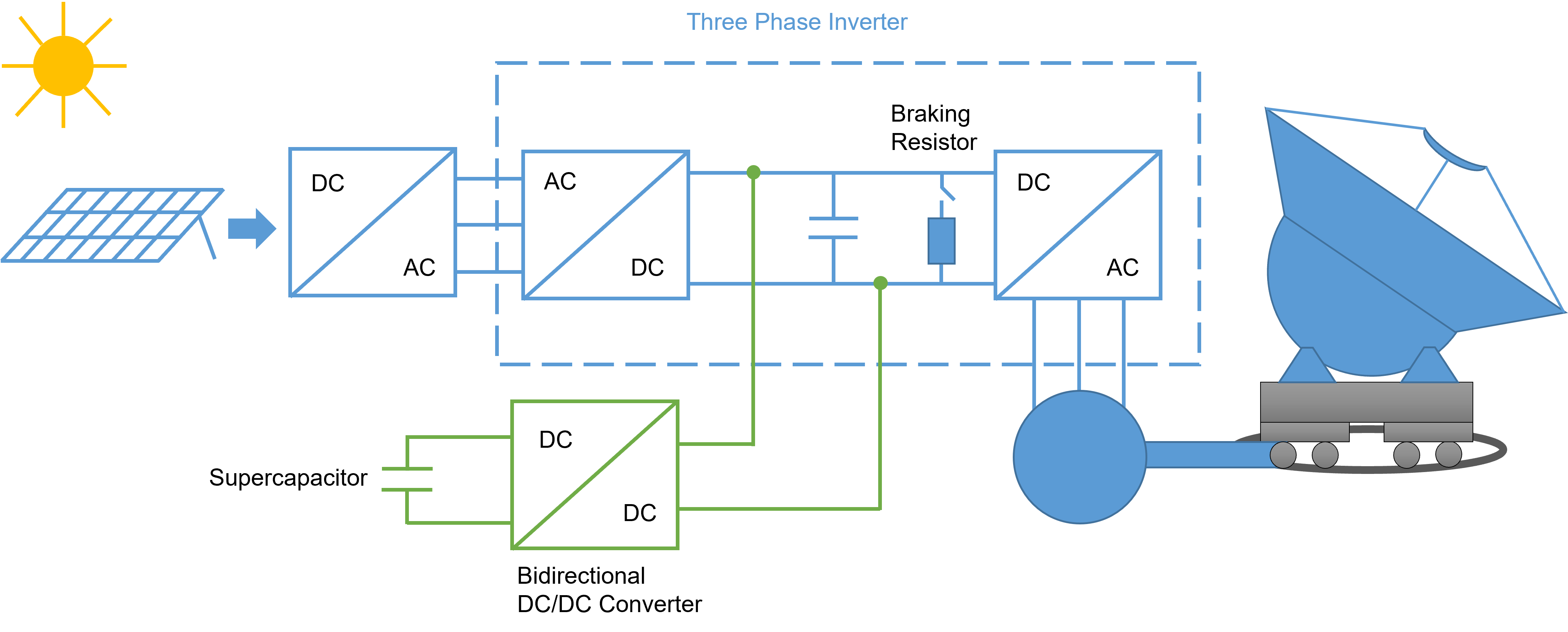}\vspace{0.2cm}
    \caption{Proposed energy recovery system.}
    \label{fig:PowerElectronicsAZ_modified}
\end{figure}

Next, the question of supercaps connection to the DC-link is addressed. In principle, a direct parallel connection of supercaps to the DC-link is possible. However, in that case the power peak shaving feature is strongly limited, as it will be shown in simulation results. Instead, we involve a bidirectional DC-DC converter between the DC-link and supercaps to gain the full control of the energy flow, as it has been previously introduced for elevator applications\cite{Jabbour}. The proposed power electronics scheme is shown in Fig.~\ref{fig:PowerElectronicsAZ_modified}.

\section{Main Motion Scenarios}\label{sec:Motion_Scenarios}
\begin{figure}[ht]
    \centering
    \begin{subfigure}{0.535\textwidth}
    \centering
    \includegraphics[width=\textwidth]{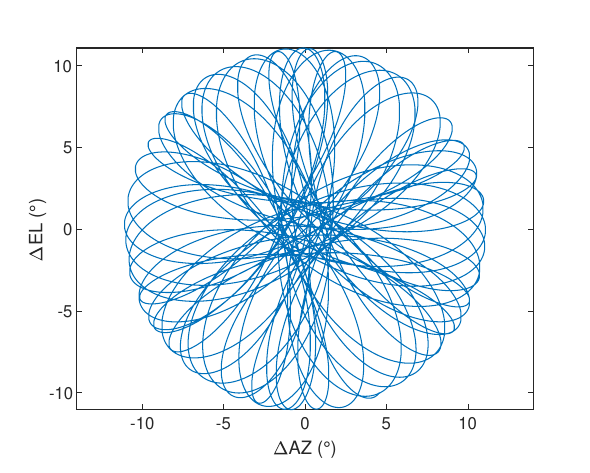}\vspace{0.5cm}
    \end{subfigure}
    \hfill
    \begin{subfigure}{0.455\textwidth}
    \centering
    \includegraphics[width=\textwidth]{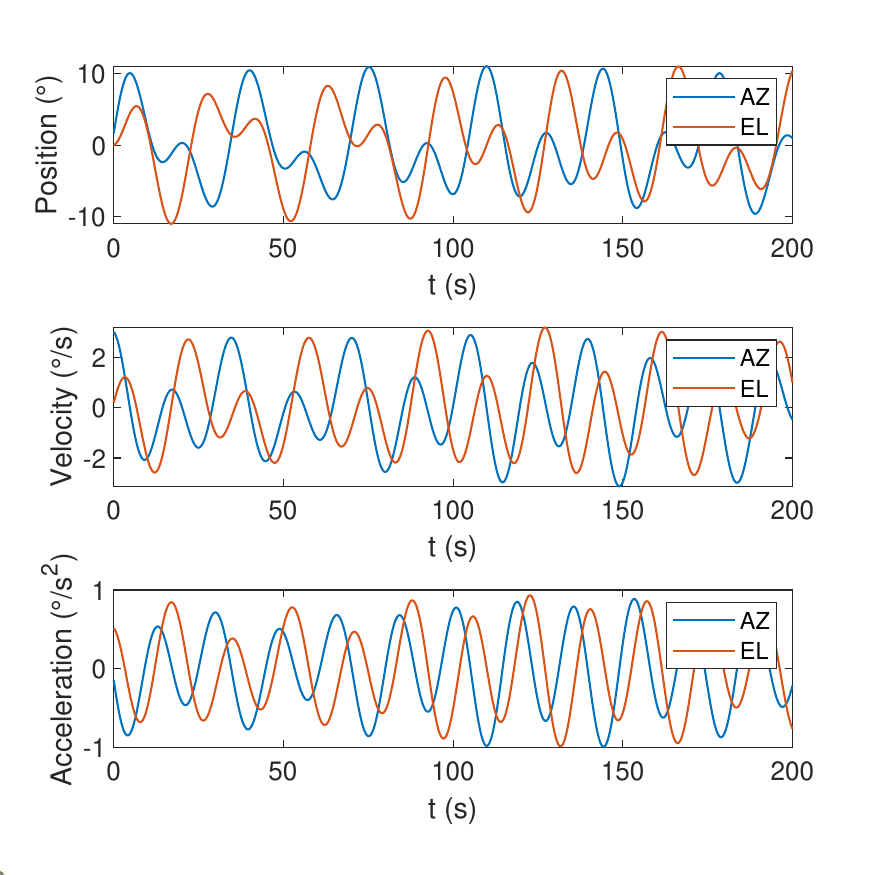}
    \end{subfigure}\vspace{0.2cm}
    \caption{Representative Lissajous daisy scan planned for AtLAST.}
    \label{fig:LissajousDaisy}
\end{figure}

The steerable axes of a telescope must not only compensate for the Earth's rotation (i.e. tracking of celestial objects), but also ensure a fast observing source change (slewing) and, often, to allow one to scan the sky rapidly (i.e. a scanning mode). Both of the latter examples are characterized by high acceleration phases, followed by a high deceleration phase (sometimes, a maximum speed phase is in between). For AtLAST, scanning is one of the key motion scenarios, with the AZ- and EL-velocity of up to $3^\circ \rm s^{-1}$ and acceleration of up to $1^\circ \rm s^{-2}$. A representative Lissajous daisy scan for AtLAST, which attains the maximum acceleration and speed, is shown Fig.~\ref{fig:LissajousDaisy} \cite{2024arXiv240218645M}.
We note that constant elevation scans frequently used in astronomical surveys present nearly the same level of acceleration and deceleration demand, but only for the azimuthal drives; the system we present here is equally suitable in this scenario. 

\section{Optimal Energy Recovery System Design}\label{sec:SystemDesign}
The proper sizing of the supercapacitors and DC-DC converters to be incorporated is crucial for the optimal cost-benefit ratio of the proposed system. The key point in this analysis is the correct calculation of the energy storage to be provided by supercaps. During deceleration, the generated power $P_g$ of a telescope's axis (neglecting wind) is due to inertia and can be expressed as follows:
\begin{equation}
    P_g=(T_I-T_f) \omega_m = (I\dot{\omega}_m-T_f) \omega_m
\end{equation}
where $I$ is the inertia of the axis and $T_f$ the friction torque. The total recuperated energy $E_g$, therefore, yields to
\begin{equation}
    E_g=\eta P_g t_{dec}
\end{equation}
where $\eta$ is the drive efficiency and $t_{dec}$ the deceleration time. Hence, the maximum recuperated energy is generated in the motion scenario of decelerating the axis from maximum velocity to zero with maximum deceleration. Applying these considerations to AtLAST, the total energy storage demand for AZ- and EL-axis combined results to $\approx 1900~\rm kW~s$ (i.e.\ 1.9~MJ). It should be noted that due to the wind on the telescope's axes, the real recuperated energy will sometimes be higher or lower, depending on the wind direction and loading (i.e. for an enclosed or domeless structure). Nevertheless, the proposed calculation covers the majority of possible scenarios, avoiding an expensive over-dimension of the system.

In the next step, the optimal supercapacitors combination must be found. A bidirectional DC-DC converter typically has one low-voltage and one high-voltage side. This fact results into the necessity of keeping the supercap voltage always either below or above the DC-link voltage. From the efficiency perspective, the voltage should be kept as high as possible to reduce the current and thus, the losses of the system. Therefore, the DC-link is set on the low-voltage side of the DC-DC converter and supercaps on the high-voltage side, respectively. This defines the minimum allowed voltage on supercaps $V_{min}$ equal to the typical maximum DC-link voltage of $660$~V. The highest supercaps voltage $V_{max}$, on the contrary, is given by the required energy storage demand and available capacitance:
\begin{equation}
    E=\dfrac{1}{2}C(V_{max}^2-V_{min}^2)
\end{equation}
For AtLAST, $V_{max}$ is achieved by a series connection of six supercaps (representing one string), each with a rated voltage of 162~V, resulting in a total voltage of $V_{max}=6\cdot162~\rm V=972$~V. Thus, to store the recovered energy a minimum system capacitance of $7.5$~F is required. However, a higher capacitance is recommended to fully leverage the peak shaving capability of the proposed system. Given also the early stage of the project, the proposed configuration has a system capacitance of 31~F, achieved by using six 92~F supercapacitors in series across two parallel strings, resulting in a total of twelve supercaps and providing a reasonable margin. As the AtLAST design matures, simulations should review the possibility of reducing the configuration to a single string of supercaps with a capacitance of $15.3$~F, respectively.

\section{Simulation Results}\label{sec:SimResults}
To verify the proposed energy recovery system, both the AZ and EL axis of AtLAST have been simulated in MathWorks Simscape.\footnote{\href{https://www.mathworks.com/products/simscape.html}{https://www.mathworks.com/products/simscape.html}} In the simulation, the inertia of the mount, friction and wind load are considered. As motion profile, Lissajous daisy scan as shown in Fig.~\ref{fig:LissajousDaisy} is commanded. The considered simulation parameters are shown in Table~\ref{tab:SimParam}.
\renewcommand{\arraystretch}{1.2}
\begin{table}[h]
    \caption{Main simulation parameters.}\vspace{0.2cm}
    \label{tab:SimParam}
    \centering
    \begin{tabular}{|p{0.5\textwidth}|p{0.2\textwidth}|}
        \hline
        Total drive efficiency & $85\%$\\
        \hline
        Capacitance of the proposed supercaps configuration & $31~\rm F$\\
        \hline
        Max.\ supercaps voltage  & $972~\rm V$\\
        \hline
        Min.\ supercaps voltage  & $660~\rm V$\\
        \hline
        Wind speed  & $10~\rm m/s$\\
        \hline
    \end{tabular}
\end{table}

\subsection{Classic Power Electronics Scheme}
\begin{figure}[ht]
    \centering
    \includegraphics[width=0.95\textwidth]{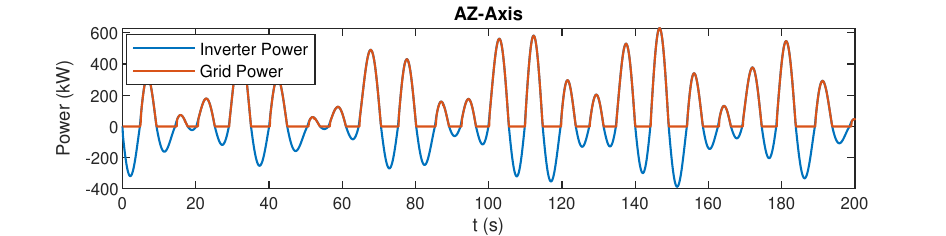}
    \includegraphics[width=0.95\textwidth]{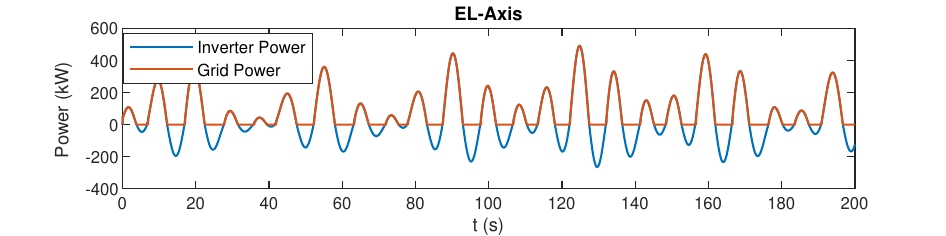}\vspace{0.2cm}
    \caption{Electric power to perform the representative Lissajous daisy scan vs.\ corresponding grid power for AZ- and EL-axis. The area within the negative values of the inverter power curve (blue) is the energy that can be recuperated.}
    \label{fig:Power_No_Supercaps}
\end{figure}

First, the classic power electronics scheme from Fig.~\ref{fig:PowerElectronicsAZ} is simulated. The grid power and the inverter power required to perform the scan are presented in Fig.~\ref{fig:Power_No_Supercaps}. 
As expected, the grid covers the total power demand when power is consumed, and blocks when power is generated. However, the power demand of the AZ-axis is mostly shifted w. r. t. the EL-axis power demand. Thus, there are time periods, where the drives of one axis consume power, while the other axis generates it. Utilizing this effect would decrease the grid power load of the telescope's main drives at no cost, since the energy exchange between the axes can be easily achieved by connecting their DC-links. 

\begin{figure}[ht]
    \centering
    \includegraphics[width=1\textwidth]{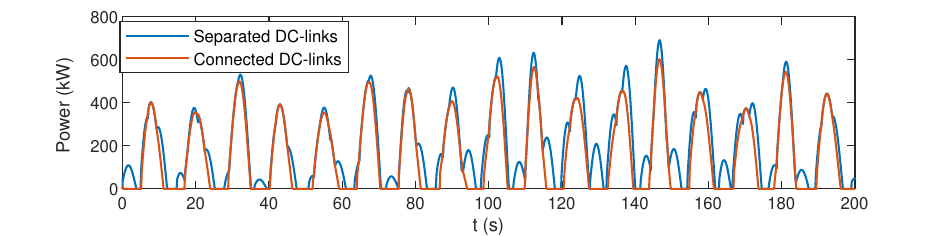}\vspace{0.2cm}
    \caption{Total power grid demand with separated vs.\ connected AZ/EL DC-links.}
    \label{fig:Power_No_Supercaps_DCVerbund}
\end{figure}

The DC-link connection is therefore useful and is foreseen for AtLAST. The total grid power with separated vs.\ connected DC-links is shown in Fig.~\ref{fig:Power_No_Supercaps_DCVerbund}. For the considered Lissajous daisy scan, this measure alone decreases the RMS power demand to the grid by $11\%$.

\subsection{Extension by Supercaps without DC-DC Converter}
\begin{figure}[ht]
    \centering
    \includegraphics[width=1\textwidth]{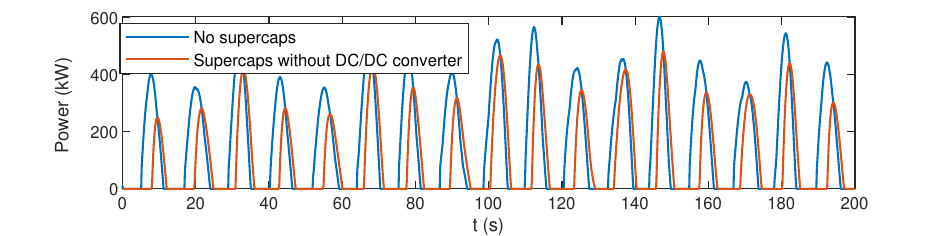}
    \includegraphics[width=1\textwidth]{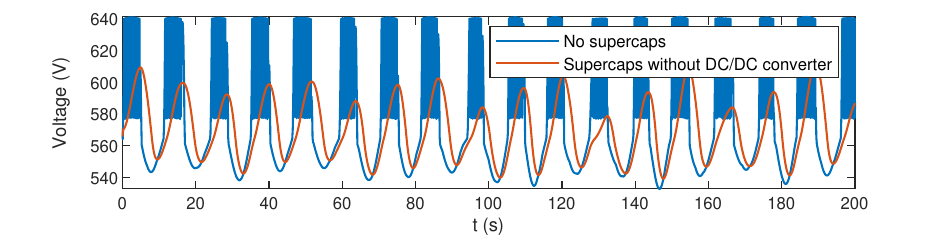}\vspace{0.2cm}
    \caption{Grid power (top) and DC-link voltage (bottom) without supercaps vs.\ with directly connected supercaps (no DC-DC converter).}
    \label{fig:Power_Supercaps_NoDCDC_Converter}
\end{figure}

Next, the DC-link of each axis is directly connected to the described supercaps configuration (six serial supercaps, two parallel strings) and thus, interconnected. The power provided by the grid and the DC-link voltage are shown in Fig.~\ref{fig:Power_Supercaps_NoDCDC_Converter}. 
From the figure it becomes clear, that the recuperated energy charges the supercaps without overloading them, as is the case at the stand-alone DC-link capacitor (for comparison, we note the two-level control behavior of the DC-link voltage without supercaps in the lower panel of Fig.~\ref{fig:Power_Supercaps_NoDCDC_Converter}). Thus, the recuperated energy is entirely recoverable. However, the grid power peaks are still high, since supercaps are passive and only follow the power dictated by the drives. 

\subsection{Full Energy Recovery System}
\begin{figure}[ht]
    \centering
    \includegraphics[width=1\textwidth]{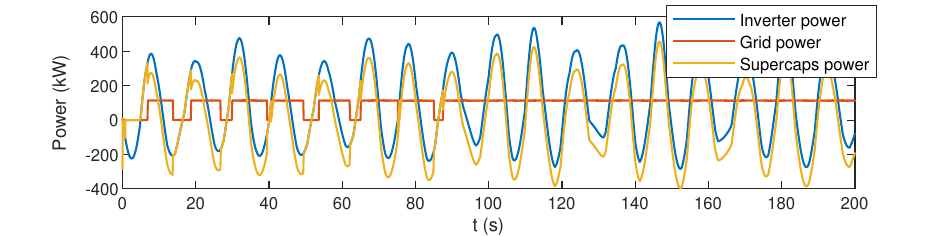}
    \includegraphics[width=1\textwidth]{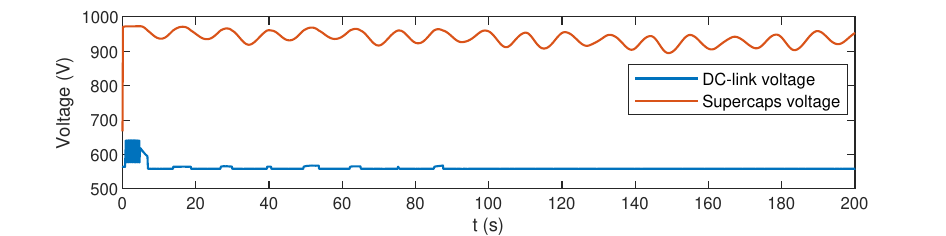}\vspace{0.2cm}
    \caption{Electric power from inverter vs.\ grid vs.\ supercaps perspective (top) and voltage at DC-link vs.\ supercaps (bottom).}
    \label{fig:Power_Voltage_Supercaps}
\end{figure}

Finally, the proposed power electronics scheme with six serial supercapacitors and controlled DC-DC converter is simulated. The DC-DC converter operates in current control mode, where the reference value is the current flowing between the DC-link and the motors. An additional offset on the reference is imposed to refill the energy loss due to friction and drive efficiency. The results are shown in Fig.~\ref{fig:Power_Voltage_Supercaps}. 
As is evident from the figure, the use of the DC-DC converter allows a full control of supercaps charge/discharge process. During acceleration, drives are powered mostly by supercaps and supported by the grid. During deceleration, supercaps are charged mostly by the drives and again, supported by the grid. This strategy effectively shaves the power peaks.

\begin{figure}[ht]
    \centering
    \includegraphics[width=1\textwidth]{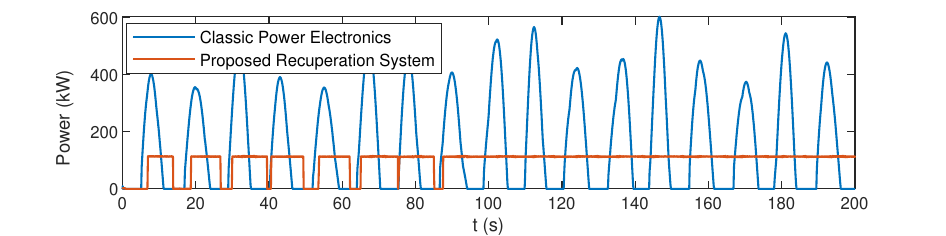}\vspace{0.2cm}
    \caption{Grid power load of the classic vs.\ proposed power electronics scheme.}
    \label{fig:Power_Supercaps_vs_classic}
\end{figure}

To summarize the efficiency improvement and grid power peak shaving impact, the grid power demand of the classic power electronics scheme is directly compared to the results of the proposed system in Fig.~\ref{fig:Power_Supercaps_vs_classic}. The RMS power demand to the grid for the considered representative Lissajous daisy scan could be decreased by $56\%$, while the power peaks were reduced by $80\%$. As a result, this approach not only halves the energy required for telescope's motion but also significantly reduces the demand on the grid infrastructure.

\section{Conclusion}
The strategy of recuperating kinetic energy to enhance motion efficiency is well-established in various applications. However, to the best of our knowledge, it has not been previously implemented in large telescopes. The latter is distinct from most other applications due to the combination of moderate energy levels and very high power peaks that must be managed. In this work we analyzed a solution based on supercapacitors, which easily serve the power peaks expected, combined with a DC-DC converter to fully control the charge/discharge process of the supercaps, w.r.t.\ large telescopes. The presented system is applied to the planned Atacama Large Aperture Submm Telescope. The simulation results confirm a remarkable decrease in power demand for the telescope's drives as the largest power consumer, halving the RMS motion power required from the grid and reducing the grid power peaks by $80\%$. 

\acknowledgments 

This project has received funding from the European Union’s Horizon 2020 research and innovation program under grant agreement No.\ 951815 (AtLAST).\footnote{See
\href{https://cordis.europa.eu/project/id/951815}{https://cordis.europa.eu/project/id/951815}.}

\bibliography{report} 
\bibliographystyle{spiebib} 

\end{document}